\documentclass[twocolumn,eqsecnum,aps,prb,showpacs]{revtex4}

\usepackage{graphicx}

\begin{document}   

\title{Static structure factor for 
graphene in a magnetic field}
\author{K. Shizuya}
\affiliation{Yukawa Institute for Theoretical Physics\\
Kyoto University,~Kyoto 606-8502,~Japan }

\begin{abstract} 
A close study is made of the static structure factor for
graphene in a magnetic field at integer filling factors $\nu$,
with focus on revealing possible signatures of 
$\lq\lq$relativistic" quantum field theory in the low-energy physics of graphene.
It is pointed out, in particular, that for graphene even the vacuum state has 
a nonzero density spectral weight, which, together with the structure factor 
for all $\nu$, grows significantly with increasing wave vector;
such unusual features of density correlations are a "relativistic" effect 
deriving from massless Dirac quasiparticles in graphene. 
Remarkably it turns out that the  zero-energy Landau levels of electrons or holes, 
characteristic to graphene, remain indistinguishable 
in density response from the vacuum state,
although they are distinct in Hall conductance. 
\end{abstract}


\maketitle

\section{Introduction}

 A great deal of attention has recently been directed to graphene, 
a monolayer of carbon atoms, both experimentally~\cite{NG,ZTSK,ZJS} and
theoretically~\cite{ZA,Massgap,GS,PGN,NM,AF,FL}.
Graphene is marked with its novel charge carriers that behave like
massless Dirac fermions with effective speed of light 
$v_{\rm F} \approx 10^{6}$~m/s $\approx c/300$.
It thus provides a special opportunity to study $\lq\lq$relativistic" quantum dynamics 
in condensed-matter systems. 
Experiments have revealed a number of exotic transport properties of
graphene, such as the half-integer quantum Hall (QH) effect and minimal conductivity.

The dynamics of Dirac fermions becomes particularly interesting in a magnetic field
and leads to peculiar quantum phenomena, 
such as fermion number fractionalization~\cite{J} 
and spectral asymmetry,~\cite{NS,Hal,FScs}  
intimately tied to the chiral anomaly in 1+1 dimensions. 
Actually the half-integer QH effect and the presence of the zero-energy Landau levels 
observed~\cite{NG,ZTSK} in graphene are a manifestation of
fermion number fractionalization.

It would be important to explore further possible signatures of
relativistic quantum field theory in the low-energy physics of graphene.
An interesting proposal~\cite{klein} along this direction is to simulate 
the Klein paradox~\cite{kleinparadox} (or tunneling) in graphene.
Calculations~\cite{Ando} of the dielectric function also reveal 
that the electromagnetic response of graphene is substantially different 
from that of conventional two-dimensional systems.
The difference becomes even prominent under a strong magnetic field.~\cite{KSgr}
In particular, for graphene the vacuum state is 
a dielectric medium and carries an appreciable amount of 
electric and magnetic susceptibilities over all range of wavelengths;
this reflects the presence of the $\lq\lq$Dirac sea".
Curiously the zero-energy Landau levels, 
though distinct in Hall conductance, hardly contribute to the susceptibilities.

The purpose of this paper is to study further aspects of 
the response of graphene in a magnetic field at integer filling factors $\nu$,
with focus on the static structure factor 
$s({\bf p}) \sim \langle \rho_{\bf -p} \rho_{\bf p}\rangle$,
which is directly related to the cross section 
in inelastic light scattering by graphene. 
In particular, we point out that graphene has unusual characteristics of 
electronic correlations at short distances, which derive 
from the $\lq\lq$relativistic" nature of massless quasiparticles. 
It is also shown that the zero-energy Landau levels remain indistinguishable
in spectral weight $\langle \rho_{\bf -p} \rho_{\bf p}\rangle$ from the vacuum state.

In Sec.~II we briefly review the low-energy     effective theory 
of graphene in a magnetic field and derive, as a preliminary, 
the structure factor for conventional QH systems.
In Sec.~III we study the structure factor for graphene.
Section~IV is devoted to a summary and discussion.

\section{Low-energy effective theory}

Graphene has a honeycomb lattice consisting of two triangle
sublattices of carbon atoms with one electron per site.
It is a gapless semiconductor and its low-energy electronic transport 
is described by an effective Hamiltonian of the form~\cite{Semenoff}
\begin{eqnarray}
H&=& \int d^{2}{\bf x}\Big[ \psi^{\dag} {\cal H}_{+} \psi 
+ \chi^{\dag} {\cal H}_{-}\chi \Big], \nonumber\\
{\cal H}_{\pm}&=& v_{\rm F}\,  (\sigma_{1}\Pi_{1} 
+ \sigma_{2}\Pi_{2} \pm m \sigma_{3})  - e A_{0}, 
\label{Hzero}
\end{eqnarray}
where $\Pi_{i} = -i\partial_{i} + e A_{i}$ [$i= (1,2)$ or $(x,y)$] 
includes coupling to external electromagnetic potentials
$A_{\mu}= (A_{i}, A_{0})$;
$v_{\rm F} \sim 10^{6}$ m/s is the Fermi velocity. 
The two-component spinors $\psi= (\psi_{1}, \psi_{2})^{\rm t} $ and 
$\chi= (\chi_{1}, \chi_{2})^{\rm t} $ stand for the electron fields
near the two inequivalent Fermi points ($K$ and $K'$)  
where the spectrum becomes linear;
$(\psi_{1} , \chi_{2})$ reside on the same sublattice and 
$(\psi_{2}, \chi_{1})$ on another.

For generality we have introduced a tiny $\lq\lq$mass" gap $m >0$
which spoils the valley SU(2) symmetry of $H$.
Actually the observed $\nu=\pm 1$ Hall plateaus~\cite{ZJS} suggest 
such a tiny mass gap.~\cite{NM,AF,FL} 
We keep $m\not=0$ to clarify the particle-hole character of 
the lowest Landau levels, but practically set $m\rightarrow 0$.

We suppress the electron spin, which is treated
as a global SU(2) symmetry of $H$ by doubling the fields
$\psi$ and $\chi$. 
The Zeeman splitting, though ignored for simplicity, 
is readily incorporated.

The Coulomb interaction is written as
\begin{equation}
H^{\rm Coul} 
= {1\over{2}} \sum_{\bf p}
v_{\bf p}\, \rho_{\bf -p}\, \rho_{\bf p},
\label{Hcoul}
\end{equation}
where $\rho_{\bf p}$ is the Fourier transform of the electron number
density $\rho = \psi^{\dag}\psi + \chi^{\dag}\chi$;
$v_{\bf p}= 2\pi \alpha/(\epsilon_{\rm b} |{\bf p}|)$ is 
the Coulomb potential with the fine-structure constant 
$\alpha = e^{2}/(4 \pi \epsilon_{0}) \approx 1/137$ and 
the substrate dielectric constant $\epsilon_{\rm b}$.  
We shall discuss the effect of $H^{\rm Coul}$ later.

Let us place graphene in a strong magnetic field and 
study how the electrons in graphene respond 
to a weak potential $A_{0}(x)$. 
We set $A_{i}\rightarrow B\, (-y,0)$ 
to supply a uniform magnetic field $B_{z}=B>0$ 
normal to the sample plane.

When $A_{0}=0$, the eigenmodes of $H$ are
Landau levels of $\psi$ and $\chi$ of energy 
\begin{equation}
\epsilon_{n} = s_{n}\,  \omega_{\rm c} 
\sqrt{|n| + m^{2}\ell^{2}/2},
\end{equation}
labeled by integers $n=0,\pm 1, \pm2, \dots$, and
$p_{x}$ (or $y_{0} \equiv \ell^{2} p_{x}$ with the magnetic length 
$\ell \equiv 1/\sqrt{eB}$); 
$\omega_{\rm c} = \sqrt{2}\, v_{\rm F}/\ell$ is the basic cyclotron
frequency.  Here $s_{n} \equiv {\rm sgn}\{n\} =\pm 1$ specifies 
the sign of the energy $\epsilon_{n}$.

For $n\not=0$, $\psi$ and $\chi$ have the same spectrum 
symmetric about $\epsilon=0$. 
The $n=0$ level of $\psi$ has negative energy 
$\epsilon_{0_{-}} = - v_{\rm F} m$ while that of $\chi$ has positive
energy $\epsilon_{0_{+}} = v_{\rm F} m$;
these $n=0_{\mp}$ levels represent holes and electrons via quantization.
With the electron spin taken into account, each Landau
level is thus four-fold degenerate, except for the doubly-degenerate
$n=0_{\pm}$ levels.
The $n=0_{\pm}$ eigenmodes have components only 
on each separate sublattice.

To make this Landau-level structure explicit, it is useful to pass to
the $|n,y_{0}\rangle$ basis, with the expansion~\cite{KSproj}
$\psi ({\bf x}, t) = \sum_{n, y_{0}} \langle {\bf x}| n, y_{0}\rangle\, 
\psi_{n}(y_{0},t)$. (From now on, we shall only display the $\psi$
sector since the $\chi$ sector is 
obtained by reversing the sign of $m$.)
The Hamiltonian $H$ thereby is rewritten as
\begin{eqnarray}
H\! &=& \!\! \int\! dy_{0} \!\!\!
\sum_{n =-\infty}^{\infty} \!\!\!
\psi^{\dag}_{n}\, \epsilon_{n}\, \psi_{n},
\label{Hzeronn}
\end{eqnarray}
and the charge density 
$\rho_{-{\bf p}}(t) =\int d^{2}{\bf x}\,  
e^{i {\bf p\cdot x}}\,\psi^{\dag}\psi$ as~\cite{KSgr}
\begin{equation}
\rho_{-{\bf p}} = e^{-\ell^{2} {\bf p}^{2}/4}
\sum_{k, n=-\infty}^{\infty} g^{\psi}_{k n}({\bf p})\int dy_{0}\,
\psi_{k}^{\dag}\, e^{i{\bf p\cdot r}}\,
\psi_{n} , 
\nonumber
\end{equation}
with the coefficient matrix
\begin{eqnarray}
g^{\psi}_{n n'}({\bf p}) &=& \textstyle{1\over{2}}\, \Big[
c_{n}^{+}\, c_{n'}^{+}\, 
f_{|n|\! -\!1, |n'| -\!1}({\bf p}) \nonumber\\
&&\ \ \  + s_{n}s_{n'}c_{n}^{-}\, c_{n'}^{-}\, 
f_{|n|, |n'|}({\bf p}) \Big] 
\label{gpsikn}
\end{eqnarray}
and $c_{n}^{\pm} = \sqrt{1 \pm  v_{\rm F} m/\epsilon_{n}}$;
${\bf r} = (r_{1}, r_{2}) = (i\ell^{2}\partial/\partial y_{0}, y_{0})$
stands for the center coordinate with uncertainty 
$[r_{1}, r_{2}] =i\ell^{2}$.
Here the coefficient functions 
\begin{eqnarray}
f_{k n}({\bf p}) 
&=& (k|e^{-{i(p/\sqrt{2})\, a^{\dag}}}\,
e^{-i(p^{\dag}/\sqrt{2})\,a}\,|n)
\end{eqnarray}
are defined in terms of the harmonic oscillator eigenstates $\{|n)\}$ with
$a^{\dag}a |n) =n|n)$ and $[a, a^{\dag}]=1$; $p=p_{y}\! +i p_{x}$
and $p^{\dag}=p_{y}\! -i p_{x}$. 
More explicitly,~\cite{GJ}  
\begin{eqnarray}
f_{k n}({\bf p}) 
&=& \sqrt{{n!\over{k!}}}\,
\Big({i\ell p\over{\sqrt{2}}}\Big)^{k-n}\, L^{(k-n)}_{n}
\Big(\textstyle{1\over{2}} \ell^{2}{\bf p}^{2}\Big)
\end{eqnarray}
for $k \ge n$, and $f_{n k}({\bf p}) = [f_{k n}({\bf -p})]^{\dag}$.
Actually $f_{k n}({\bf p})$ are the coefficient functions that
characterize the charge density for the "nonrelativistic"  
Hall electrons
\begin{equation}
\rho_{-{\bf p}} = e^{- \ell^{2} {\bf p}^{2}/4}
\sum_{k, n =0}^{\infty} f_{k n}({\bf p})\int dy_{0}\,
\psi_{k}^{\dag}\, e^{i{\bf p\cdot r}}\, \psi_{n}
\label{rhoNR}
\end{equation}
expressed in terms of their eigenmodes $\{\psi_{n}(y_{0},t)\}$ 
with energy $\epsilon_{n}= (eB/m^{*})(n + 1/2)$.

Let us first consider, as an exercise, the static structure factor or
the spectral weight  
$\langle \rho_{\bf -p}\, \rho_{\bf p} \rangle \equiv 
\langle G| \rho_{\bf -p}\, \rho_{\bf p}|G\rangle$ for a state $|G\rangle$ 
of free "nonrelativistic" Hall electrons with integer filling factor $\nu$ (per spin).
One may use Eq.~(\ref{rhoNR}) and note, in taking the matrix element 
$\langle G| \rho_{\bf -p} \rho_{\bf p}|G\rangle$, that 
$\psi_{k'}(y'_{0},t)\psi^{\dag}_{k}(y_{0},t) \rightarrow 
\delta_{k' k} \delta (y'_{0}- y_{0})$  for unoccupied levels $(k',k)$
so that the result is proportional to the number of electrons per occupied level
$\int dy_{0}\, \psi^{\dag}_{n}\psi_{n} \rightarrow  
L_{x}L_{y}/(2\pi \ell^{2})$.  (Note here that 
$\delta (y_{0}=0) = L_{x}/(2\pi \ell^{2})$ with $L_{x} = \int dx$.)
This yields the spectral weight 
\begin{equation}
\langle \rho_{\bf -p}\, \rho_{\bf p}\rangle
= {\Omega\over{2\pi \ell^{2}}}\,  e^{- \ell^{2}
{\bf p}^{2}/2} \sum_{k=\nu}^{\infty}\sum_{n=0}^{\nu-1} |f_{kn}({\bf p})|^{2}
\label{rhorhoNR}
\end{equation}
for ${\bf p}\not= 0$, where $\Omega = L_{x}L_{y}$ denotes the total area.

Let us recall that the static structure factor~\cite{mahan}
\begin{equation}
s({\bf p})= (\langle \rho_{\bf -p}\, \rho_{\bf p}\rangle
 -N_{\rm e}^{2} \delta_{{\bf p}, 0})/N_{\rm e} 
\end{equation}
is defined by $\langle \rho_{\bf -p}\, \rho_{\bf p}\rangle$ 
with its ${\bf p}=0$ component isolated, where 
$N_{\rm e} = \nu\,\Omega/(2\pi \ell^{2})$
stands for the total electron number.
Accordingly, in general,   $s({\bf p}\rightarrow 0) =0$ owing to charge conservation.
Noting the formula 
\begin{equation}
\sum_{k=0}^{\infty} |f_{kn}({\bf p})|^{2} = e^{ \ell^{2}{\bf p}^{2}/2}  
\label{completeNR}
\end{equation}
allows one to cast Eq.~(\ref{rhorhoNR}) into the structure factor 
at filling factor $\nu$ 
\begin{equation}
s({\bf p})
= 1 - e^{- \ell^{2} {\bf p}^{2}/2}\, {1\over{\nu}}\,  
\sum_{n=0}^{\nu-1}  \sum_{k=0}^{\nu-1} |f_{kn}({\bf p})|^{2},
\label{spNR}
\end{equation}
which agrees with a known result.~\cite{REF}
In particular, 
\begin{equation}
s({\bf p})|_{\nu=1} = 1 -  e^{- \ell^{2} {\bf p}^{2}/2}.
\end{equation}

Note first that  $s({\bf p}\rightarrow 0) \rightarrow  0$
since $f_{kn}(0) = \delta_{kn}$. 
Note also  that $s({\bf p}) \rightarrow 1$
as ${\bf p}\rightarrow \infty$ for all $\nu$.  
To see what this means let us recall the following: 
For a collection of classical particles, $s({\bf p})$ is written as~\cite{mahan}
\begin{equation}
s({\bf p}) -1 = \Big\langle \sum_{\triangle {\bf r}}e^{i {\bf p \cdot \triangle r}}
\Big\rangle
\end{equation}
for ${\bf p}\not=0$, i.e., as an average  
over the relative positions ${\bf \triangle r}$ of 
particles surrounding a given particle.  As a result, 
$s({\bf p}) \rightarrow 1$ for ${\bf p}\rightarrow \infty$ 
if $\triangle {\bf r}\not=0$,
e.g., for particles formulated on a lattice.
Such behavior of  $s({\bf p})$ thus implies the absence of particle correlations 
at short distances.

\section{Static response of graphene} 

In this section we study the case of graphene. In the present treatment 
the $\psi$ and $\chi$ sectors are
independent and the spectral weight $\langle \rho \rho \rangle$ 
with $\rho= \rho^{\psi} + \rho^{\chi}$ 
is given by the sum 
$\langle \rho^{\psi} \rho^{\psi} \rangle 
+ \langle \rho^{\chi} \rho^{\chi} \rangle$ for ${\bf p}\not=0$.
For the $\psi$ sector one may
simply replace, in Eq.~(\ref{rhorhoNR}),
$f_{kn}({\bf p})$ by $g^{\psi}_{kn}({\bf p})$ of Eq.~(\ref{gpsikn})
and note that the level indices $(k,n)$ now run over all integers, 
$0_{-}, \pm 1, \pm 2, \cdots$. 
Analogously, $g^{\chi}_{kn}({\bf p})$ for the $\chi$ sector 
is obtained from $g^{\psi}_{kn}({\bf p})$ by setting
$m \rightarrow -m$ (i.e., $c^{+}_{n}\leftrightarrow c^{-}_{n}$), or equivalently
\begin{equation}
g^{\chi}_{kn}({\bf p}) = g^{\psi}_{-k,-n}({\bf p}). 
\label{gpsigchi}
\end{equation}
Let us denote $\Gamma^{\psi}_{k n}(z) 
= |g^{\psi}_{k n}({\bf p})|^{2}$ and $\Gamma^{\chi}_{k n}(z) 
= |g^{\chi}_{k n}({\bf p})|^{2}$ for short.  They actually are
functions of $z=\ell^{2}{\bf p}^{2}/2$, are thus symmetric in 
$(k,n)$, and have the property 
\begin{equation}
\Gamma^{\psi}_{k n} = \Gamma^{\chi}_{-k, -n},  
\Gamma^{i}_{k n} = \Gamma^{i}_{n k},   
\label{Gammapsichi}
\end{equation}
with $i=\psi,\chi$.
The spectral weight is now written as
\begin{equation}
\langle \rho_{\bf -p} \rho_{\bf p}\rangle / \Omega
= (\lambda_{\rm s}/2\pi \ell^{2})\, 
e^{-{1\over{2}}\,\ell^{2} {\bf p}^{2}} (I^{\psi} + I^{\chi} ),
\end{equation}
where $I^{i}=\sum'_{k}\sum_{n} \Gamma^{i}_{kn}(z)$; 
$\lambda_{\rm s}=2$ stands for the spin degeneracy.
One can write $I^{\psi}$ as
\begin{equation}
I_{j}^{\psi}(z) = \sum_{k>j}\sum_{n\le j} \Gamma^{\psi}_{k n}(z),
\end{equation}
when the $\psi$ Landau levels are occupied up 
to the $j$th level $(j=0_{-}, \pm 1, \dots)$;
analogously for $I_{j}^{\chi}$.

Note that Eq.~(\ref{gpsigchi}) relates $I^{\psi}$ and $I^{\chi}$ so that 
\begin{equation}
I_{j}^{\psi} =I_{-(j+1)}^{\chi}.
\nonumber
\end{equation}
The equalities $I_{0_{-}}^{\psi} =I_{-1}^{\chi}$ 
and $I_{j}^{\psi} +I_{j}^{\chi} =I_{-(j+1)}^{\psi}
+I_{-(j+1)}^{\chi}$ then imply the following:
(i)~The $\psi$ and $\chi$ sectors contribute equally to the vacuum spectral weight, 
$\langle \rho_{\bf -p}\, \rho_{\bf p}\rangle|_{\nu=0} \propto
I_{0_{-}}^{\psi} +I_{-1}^{\chi}= 2\, I_{0_{-}}^{\psi}$.
(ii)~The spectral weight 
$\langle \rho_{\bf -p} \rho_{\bf p}\rangle$ is the same for the
charge-conjugate states with $\nu=\pm 2(2j+1)$.
In view of this we shall focus on the case $\nu \ge 0$ from now on.

In the limit $m\rightarrow 0$ of practical interest, 
distinctions between $\Gamma^{\psi}$ and $\Gamma^{\chi}$ 
disappear: $\Gamma^{\psi}_{k n} =
\Gamma^{\chi}_{k n} =\Gamma^{\chi}_{-k, -n}$ for $kn \not=0$ 
and $\Gamma^{\psi}_{k, 0_{-}} =\Gamma^{\chi}_{\pm k, 0_{+}}$ for
$k\not=0$, as seen from Eq.~(\ref{gpsikn}). 
This yields 
\begin{equation}
I_{j}^{\psi} =I_{j}^{\chi}= I_{-(j+1)}^{\psi}= I_{-(j+1)}^{\chi}
\ \ (m\rightarrow 0).
\end{equation}
In particular, $I_{0_{-}}^{\psi} = I_{0_{+}}^{\chi}=I_{-1}^{\chi} =I_{-1}^{\psi}$
shows that the graphene vacuum and the $\nu=\pm 2$ states, all of zero energy,
have the same spectral weight 
$\langle \rho_{\bf -p} \rho_{\bf p}\rangle|_{\nu=2} 
\propto I_{0_{-}}^{\psi} +  I_{0_{+}}^{\chi} = 2 I_{0_{-}}^{\psi}$.
Actually we earlier noted such degeneracy in electric susceptibility 
for the  $\nu=0$ and $\nu=\pm 2$ states,~\cite{KSgr} and 
these states now turn out to be indistinguishable 
in both the real and imaginary parts of density response;
we discuss this point in more detail later.

%%%%%%%%%%%% Fig. 1 %%%%%%%%%%%%  
\begin{figure}[tbp]
\begin{center}
\scalebox{1}{
\includegraphics{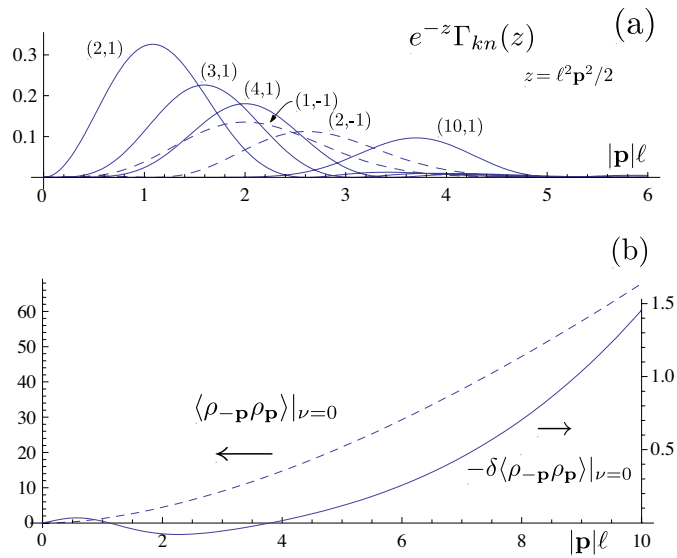}}
\end{center}
\caption{ (a) Profile of the transition rate 
$e^{-z}\Gamma_{kn}^{\psi}(z)$ for $m\rightarrow 0$, labeled 
with $(k,n)$.
(b)~Vacuum spectral weight 
$\langle \rho_{\bf -p} \rho_{\bf p}\rangle|_{\nu=0}$  for $N=800$ 
(dashed line) and the difference 
$- \delta \langle \rho_{-{\bf p}} \rho_{{\bf p}}\rangle|_{\nu=0}
=\langle \rho \rho \rangle^{B=0} - \langle \rho \rho \rangle^{B \not=0}$
(real line), both in units of 
$\lambda_{\rm s}\Omega/(2\pi \ell^{2})$.
}
\end{figure}

%%%%%%%%%%%%

To proceed further let us note some basic properties of 
the transition rates $\Gamma^{i}_{kn}(z)$  with $z = \ell^{2} {\bf p}^{2}/2$.
It is clear from Eq.~(\ref{Gammapsichi}) that 
the transition rates among the positive-energy states 
and those among the negative-energy states are essentially the same
whereas they are different from the rates across the Dirac sea
($\propto \Gamma_{kn}$ with $kn<0$).
This feature is seen from Fig.~1~(a), which illustrates typical profiles of 
$e^{-z}\Gamma_{kn}(z)$ for $m\rightarrow~0$.
In general, $e^{-z}\Gamma^{i}_{kn}(z)$ is significantly peaked 
at some value of $\ell |{\bf p}|$ and 
the peak position rises only gradually as the level gap $\sim |k-n|$ increases.
The structure factor $s({\bf p})$ for small $|{\bf p}|$ is thus governed by 
virtual transitions to neighboring levels while its property 
at larger  $|{\bf p}|$ is determined by transitions across larger gaps.

It is readily expected from this property that the sum over an infinite number 
of negative-energy levels would make the weight 
$\langle \rho_{\bf -p} \rho_{\bf p}\rangle$ cutoff ($N$) dependent 
for small ${\bf p}^{2}$, actually to $O(\ell^{2}{\bf p}^{2})$.
In particular, the vacuum weight 
$\langle \rho_{-{\bf p}} \rho_{{\bf p}}\rangle|_{\nu=0} \propto 
2 e^{-z} \sum_{k=1}^{N}\sum_{n=0}^{N} \Gamma^{\psi}_{k, -n}(z)$ 
vanishes at $z=0$ and grows rapidly 
with $z$ for fixed $N$, whereas it diverges as $N\rightarrow \infty$ for $z\not=0$.
See Fig.~1~(b). 
Indeed, evaluating the $O(z)$ terms in $\Gamma_{k,-n}(z)$
shows that the divergence is logarithmic in $N$,
\begin{eqnarray}
\sum_{k=1}^{N}\sum_{n=0}^{N}\Gamma_{k,-n}^{\psi}
&\stackrel{O(z)}{\approx}&{1\over{2}}
\, z \sum_{k=0}^{N-1} (\sqrt{k+1} -\sqrt{k})^{2}\nonumber\\
&\approx& (1/8)\, z \ln(c_{1}N)
\label{vacuumweight}
\end{eqnarray}
with $c_{1} \approx 54.088$ and for $m\rightarrow 0$.
This ultraviolet divergence is physical.
The infinite depth of the Dirac sea, of course, is an artifact of 
the continuum model~(\ref{Hzero}) and the cutoff scale 
$\omega_{\rm c}\sqrt{N}$ is set by the energy scale 
above which the model loses its validity.

Let us recall here that for conventional QH systems 
the charge operator trivially annihilates the vacuum,
$\rho |\nu=0\rangle =0$, and the vacuum spectral weight vanishes.
Accordingly, the nonzero vacuum weight 
$\langle \rho_{\bf -p} \rho_{\bf p}\rangle|_{\nu=0}$ 
itself is a "relativistic" signature of graphene,
\begin{equation}
\rho (x)|\nu=0\rangle_{\rm graphene} \not=0,
\end{equation}
which is a consequence of particle-hole pair creation
or the presence of the Dirac sea.

It is perfectly legitimate to consider the vacuum weight 
with such a physical cutoff $N$ (apart from its precise value).
One can equally well extract cutoff-insensitive information out of it. 
One possible way is to consider a variation of  
$\langle \rho_{\bf -p} \rho_{\bf p}\rangle|_{\nu=0}$ 
for $B \not=0$ and $B=0$.
Experimentally this means measuring the vacuum weight 
for $B \not=0$ and $B=0$ separately.

For $B=0$ the spectral weight for the graphene vacuum $|0\rangle$ 
with $m\rightarrow 0$ is written as 
\begin{equation}
{1\over{\Omega}}\, \langle 0|\rho_{-{\bf p}} \rho_{{\bf p}} |0 \rangle^{B=0}  
= \lambda_{\rm s} \lambda_{\rm v} \int {d^{2} {\bf k} \over{(2\pi)^{2}}}
{1\over{2}}\, 
(1 - \cos \theta_{\bf k+p,k}),
\end{equation}
which represents a collection of virtual transitions~\cite{Ando} from a
negative-energy electron state with momentum ${\bf k}$ to a
positive-energy state with
${\bf k+p}$ via the charge density $\rho$;
here $\theta_{\bf k+p,k}$ denotes the angle between ${\bf k}$
and ${\bf k+p}$;
$\lambda_{\rm s}=2$ and $\lambda_{\rm v}=2$ denote the spin and valley degeneracy.
This is again ultraviolet divergent.
For regularization we cut off the ${\bf k}$ integral 
at $|{\bf k}| = \Lambda$ and 
choose the "Fermi momentum" $\Lambda$ so that the Dirac sea accommodates
the same number of electrons  as in the $B\not=0$ case,
$N_{\rm sea}=\lambda_{\rm s}\lambda_{\rm v}\Lambda^{2}/(4\pi) 
\approx \lambda_{\rm s}\lambda_{\rm v}
(N+1/2)/(2\pi \ell^{2})$, i.e., $\Lambda^{2}\approx
2N/\ell^{2}$.

A direct calculation yields 
\begin{equation}
{1\over{\Omega}}\, \langle \rho_{-{\bf p}} \rho_{{\bf p}}\rangle^{B=0}  
= \lambda_{\rm s} \lambda_{\rm v} {{\bf p}^{2}\over{32\pi}}\, 
{ \log {16\Lambda^{2}\over{{\bf p}^{2}}}} .
\nonumber
\end{equation}
The cutoff $(N)$ dependence thus correctly disappears from the 
difference 
$\delta \langle \rho \rho \rangle 
\equiv  \langle \rho \rho \rangle^{B\not=0} - \langle \rho \rho \rangle^{B=0}$,
with the result 
\begin{eqnarray}
&&\delta \langle \rho_{-{\bf p}} \rho_{{\bf p}}\rangle|_{\nu=0}/\Omega
= (\lambda_{\rm s}/2\pi \ell^{2}) \, \delta I(z), \nonumber\\
&&\delta I(z) = 2e^{-z} I^{\psi}_{0_{-}}(z) - (z/4)\,   
\log (16 N/z) .
\end{eqnarray}
As seen from Fig.~1~(b), 
the vacuum weight $\langle \rho_{-{\bf p}} \rho_{{\bf p}}\rangle|_{\nu=0}$
differs only slightly for $B\not=0$ and $B=0$.

Actually for all $\nu$ the spectral weight 
$\langle \rho_{\bf -p} \rho_{\bf p}\rangle$ contains the vacuum fluctuations 
and the static structure factor $s({\bf p})$ is necessarily cutoff-dependent.
A possible cutoff-independent measure in experiment is 
to compare the spectral weights for $\nu\not=0$ and $\nu=0$ (with $B\not=0$). 
Correspondingly let us define by 
\begin{equation}
\triangle s({\bf p}) = \{\langle \rho_{\bf -p} \rho_{\bf p}\rangle
-\langle \rho_{\bf -p} \rho_{\bf p}\rangle|_{\nu=0} \}/ N_{\rm e}
\end{equation}
the structure factor with the vacuum contribution subtracted.
For the $\nu=2(2j+1)$ states (with $j=0, 1, \dots$),
$\triangle s({\bf p}) 
=  e^{-z} (\triangle I_{j}^{\psi} + \triangle I_{j}^{\chi} )/(2j+1)$
in terms of the cutoff-independent deviations 
$\triangle I^{\psi}_{j} =I^{\psi}_{j}-I^{\psi}_{0_{-}}$ and 
$\triangle I^{\chi}_{j} =I^{\chi}_{j}-I^{\chi}_{-1}$.
For $j\ge 0$ they are rewritten as
\begin{eqnarray}
\triangle I^{\chi}_{j} 
&=& \sum_{k=j+1}^{N}\sum_{n=0_{+}}^{j}\Gamma^{\chi}_{k,n} 
- \sum_{k=0_{+}}^{j}
\sum_{n=1}^{N}\Gamma^{\chi}_{k,-n}\ \
\label{dIj}  
\\
&\stackrel{m\rightarrow 0}{=}& \sum_{n=1}^{j} F_{n}(z)\, e^{z}
 - \sum_{k=1}^{j} \sum_{n=0_{+}}^{j}\Gamma^{\chi}_{k,n}(z) ;
\end{eqnarray}
analogously for $\triangle I^{\psi}_{j}$.
In reaching the second line, we have used the
formulas (valid for $m\rightarrow 0$ and $N\rightarrow \infty$):
\begin{eqnarray}
\sum_{k=1}^{N}\Gamma_{\pm k,0}(z) 
\!\! &=& {1\over{2}} \, (e^{z}-1),
\nonumber\\
\sum_{k=1}^{N}\Gamma_{k,n}(z)
&\stackrel{n\not=0}{=}&{1\over{2}}\, \big[1+ s_{n}\, F_{|n|}(z) \big]\,  e^{z}
-{1\over{4}}\,{ z^{|n|} \over{|n|!}},\ \ \ \ \ 
\label{sumGkn}
\end{eqnarray}
where $z = \ell^{2} {\bf p}^{2}/2$ and
\begin{eqnarray}
F_{n}(z) &=& e^{-z} \sum_{k=1}^{N}  {\rm Re}[ 
f_{n\! -\!1, k\! -\!1}(-{\bf p})\, f_{k, n}({\bf p})] ;
\end{eqnarray}
$F_{n}(0)=1$ and $F_{n}(z) \rightarrow 0$ as $z\rightarrow\infty$; 
the decrease is slower for larger $n$, as depicted in Fig.~2~(a).

The subtracted structure factor is now written as
\begin{equation}
\triangle s({\bf p}) = {2\over{2j+1}}\, \Big[\sum_{n=1}^{j} F_{n}(z) 
 - e^{-z}
\sum_{k=1}^{j}\sum_{n=0_{+}}^{j}\Gamma^{\chi}_{k,n}(z) \Big]
\end{equation}
for $\nu=2(2j+1)$.
In particular, $\triangle s({\bf p})|_{\nu=2}=0$ and
\begin{equation}
\triangle s({\bf p})|_{\nu=6} = {2\over{3}}\, 
\Big[ F_{1}(z) 
 - e^{-z} \big(1 -  {1\over{2}}\, z + {1\over{4}}\, z^2 \big) \Big].
\end{equation}

%%%%%%%%%%%%  Figure 2   %%%%%%%%%%%%%

\begin{figure}[tbp]
\begin{center}
\scalebox{1}{
\includegraphics{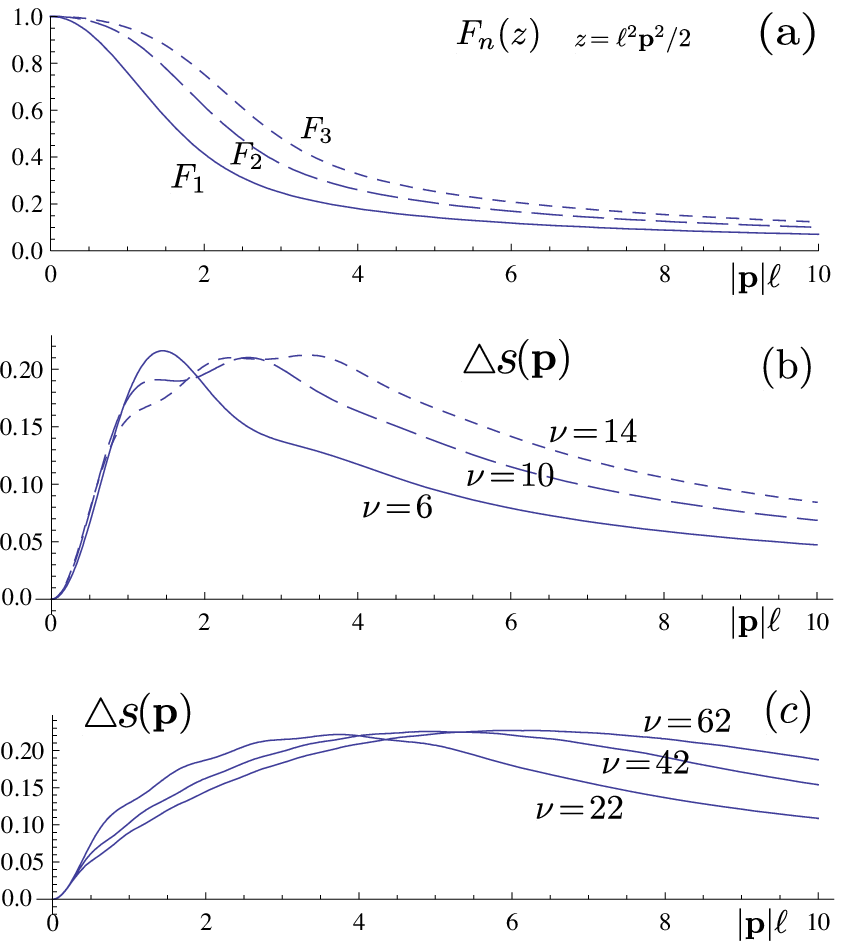}}
\end{center}
\caption{ (a) $F_{n}(z)$. (b) Subtracted static structure factor $\triangle s({\bf p})$ for $\nu$=6, 10, 14.
(c) $\triangle s({\bf p})$ for $\nu$=22, 42, 62. 
 }
\end{figure}

%%%%%%%%%%%%  Figure 3   %%%%%%%%%%%%%

\begin{figure}[tbp]
\begin{center}
\scalebox{1}{
\includegraphics{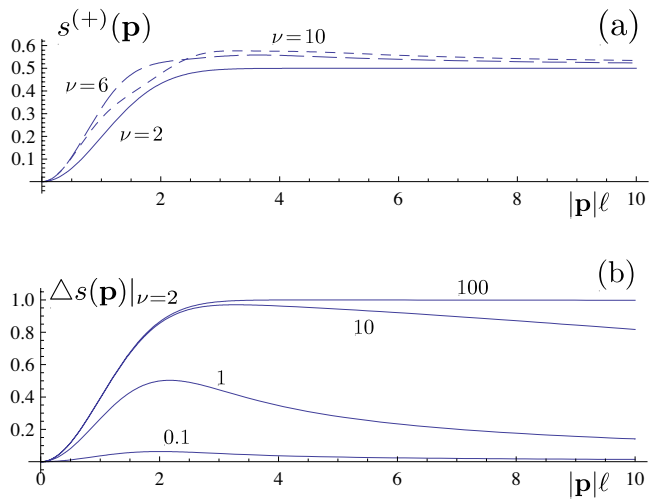}}
\end{center}
\caption{ (a) Structure factor $s^{(+)}({\bf p}$ restricted 
to the positive-energy sector, for $\nu = 2 \sim 10$.
(b)  $\triangle s({\bf p})$ at $\nu=2$ develops as the mass gap 
becomes large $v_{\rm F}m/\omega_{\rm c} = 0.1, 1, 10, 100$.
}
\end{figure}

%%%%%%%%%%%%%%%%%%%%%%%%%%%%%%

In Fig.~2 (b) and (c) we plot $\triangle s({\bf p})$ for $\nu=6, 10, 14$ and 
for higher filling $\nu=22, 42, 62$.
There are some notable features:
(1)~$\triangle s({\bf p})\rightarrow 0$ for $|{\bf p}| \rightarrow 0$.
This is a consequence of the uniformity of charge at long wavelengths.
(2)~$\triangle s({\bf p})\rightarrow 0$ for $|{\bf p}| \rightarrow  \infty$.  
This shows that the large $|{\bf p}|$ portion of $s({\bf    p})$ is common 
to all $\nu$ and is governed by 
the vacuum weight $\langle \rho_{\bf -p} \rho_{\bf p}\rangle|_{\nu=0}$.
   As a result, $s({\bf p})$ itself would rise with $|{\bf p}|$, 
in sharp contrast to the standard behavior $s({\bf p})\rightarrow 1$ 
as $|{\bf p}| \rightarrow \infty$ 
for "nonrelativistic" Hall electrons in Eq.~(\ref{spNR}). 
(3)~$\triangle s({\bf p})$ 
tends to saturate around 0.25 over a certain broad range of $|{\bf p}|$
for higher filling factors $\nu$.
The subtracted structure factor $\triangle s({\bf p})$, 
unlike $s({\bf p})$,
refers to quantum fluctuations of a finite number of filled levels above the vacuum 
and is governed by virtual transitions, newly allowed or suppressed 
by the presence of such levels, as seen from Eq.~(\ref{dIj}).
Saturation is a consequence of a competition between 
the transitions among positive-energy states and the suppressed vacuum
polarization effect.

In this connection, let us try to retain only 
the positive-energy states and transitions among them.
In Fig.~3~(a) we plot a projected structure factor $s^{(+)}({\bf p})$ 
calculated from the first term  on the right-hand side of Eq.~(\ref{dIj}).
Note that 
\begin{equation}
s^{(+)}({\bf p})\rightarrow 1/2 \ \ {\rm for} \ \  |{\bf p}| \rightarrow \infty;
\label{splusp}
\end{equation}
i.e., even the positive-energy sector alone does not recover the  
$s({\bf p})\rightarrow 1$ behavior of conventional QH systems.
While $s^{(+)}({\bf p})$ itself is not directly observable in experiment,
this feature~(\ref{splusp})  would indicate indirectly 
that  $\triangle s({\bf p})$ is significantly smaller than 1/2.

Some remarks on the influence of a mass gap are in order here.
The relativistic treatment makes sense when the mass gap is 
tiny $m \ell \ll 1$.
In the "nonrelativistic" limit  $m \ell \gg \sqrt{N}$ where 
the mass gap is large compared with the depth of the Dirac sea, in contrast,
the Landau-level sum is limited to a finite interval $N \ll m^{2}\ell^{2}$ 
and ceases to yield a divergence.
The virtual transition rates  across the mass gap scale like 
$\Gamma_{k, -n} \propto 1/(m\ell)^{2}$ 
(while $\Gamma_{kn} \sim O(1)$ for $kn>0$), and the vacuum spectral weight 
$\langle \rho_{\bf -p} \rho_{\bf p}\rangle|_{\nu=0} \propto\sum_{k,n}\Gamma_{k, -n}$
tends to zero like $(m\ell)^{-2} \log (m^{2}\ell^{2})\rightarrow 0$ as 
$m\ell \rightarrow \infty$.
Figure~3~(b) shows the manner how $\triangle s({\bf p})|_{\nu=2}$ develops 
as the mass gap gets large
 for $v_{\rm F} m/\omega_{\rm c} = 0.1 \rightarrow 100$.
The $\nu= 2$ state is now distinguishable in $s({\bf p})$ from the vacuum.
In the $m\ell \rightarrow \infty$ limit, 
$\triangle s({\bf p}) \approx s({\bf p}) \approx s^{(+)}({\bf p})$ 
recovers the standard behavior $s({\bf p})\rightarrow 1$ 
as $|{\bf p}| \rightarrow \infty$ for all integer $\nu$.

To explore the origin of the unconventional behavior~(\ref{splusp})    
let us note the formula
\begin{equation}
\sum_{k=-\infty}^{\infty} |g_{kn}({\bf p})|^{2} = e^{ \ell^{2}{\bf p}^{2}/2}. 
\label{completedirac}
\end{equation}
which follows from Eq.~(\ref{sumGkn}).
Actually this formula is valid for $m\not=0$ as well, 
since it is independently derived from the evaluation of the static weight 
$\langle G| \rho_{\bf -p} \rho_{\bf p} |G\rangle$
for a hypothetical state $|G\rangle$ with only one filled Landau level $\{n\}$ 
and all other levels empty. It is thus a consequence of completeness 
of physical states.

In the "nonrelativistic" limit 
the Dirac sea effectively disappears.
Then Eq.~(\ref{completedirac}) is reduced to Eq.~(\ref{completeNR}), 
which in turn leads to the behavior 
$s({\bf p})\rightarrow 1$ for $|{\bf p}| \rightarrow \infty$.

For $m\rightarrow 0$ the large-$|{\bf p}|$ behavior of Eq.~(\ref{completedirac})
is governed by the sums over $|k| \gg |n|$ so that 
\begin{equation}
 e^{-\ell^{2}{\bf p}^{2}/2} \sum_{k\sim0 }^{\infty} |g_{\pm k, n}({\bf p})|^{2}
 \rightarrow 1/2 , 
\label{Gammatohalf}
\end{equation}
as seen also from Eq.~(\ref{sumGkn}).
This precisely accounts for the behavior of $s^{(+)}({\bf p})$ in Eq.~(\ref{splusp}).
It is now clear that both the masslessness (or a tiny mass gap) of the quasiparticles
and the presence of the Dirac sea are crucial for the unusual density correlations 
$s^{(+)}({\bf p})\rightarrow 1/2$ at short distances.
In other words, correlations fail to vanish at short distances
because of particle-hole pair creation; the inability of localizing massless particles
is a purely  "relativistic" effect, underlying also 
the Klein paradox.~\cite{klein}

Equation~(\ref{completedirac}) may suggest that the vacuum spectral weight 
$\langle \rho_{\bf -p} \rho_{\bf p}\rangle|_{\nu=0}$ would 
have a contribution of $O(N)$ from the Dirac sea,
It actually rises like $\ln N$, as we have seen in Eq.~(\ref{vacuumweight}).
Had we defined the "vacuum" static structure factor $s({\bf p})$ with normalization 
by the total number of electrons in the Dirac sea, it would vanish like 
$s({\bf p}) \propto (\ln N)/N \rightarrow 0$.
This shows again that, for massless particles and in the presence of the Dirac sea, 
the standard (classical) picture of particle correlations 
does not necessarily make sense.

Finally we wish to comment on some peculiarities of the zero-energy $n=0_{\pm}$ levels, 
especially on how the Coulomb interaction 
$v_{\bf p}= 2\pi \alpha/(\epsilon_{\rm b} |{\bf p}|)$
affects their response.
In the random-phase approximation (RPA) the effect of 
$H^{\rm Coul}$ of Eq.~(\ref{Hcoul}) is readily included~\cite{mahan} 
into the polarization function ($\sim -i \langle \rho \rho\rangle$),
\begin{equation}
P_{\rm RPA}({\bf p},\omega) = P({\bf p},\omega)/
\{ 1 - v_{\bf p}\, P({\bf p}, \omega) \} ,
\end{equation}
once one knows the polarization function $P({\bf p}, \omega)$ 
for noninteracting Hall electrons.
Suppose now that we start with the vacuum state and fill up the $n=0_{+}$ level 
to reach the $\nu =2$ state. Then $P({\bf p}, \omega)$ will change by an amount 
proportional to 
\begin{eqnarray}
\triangle P({\bf p}, \omega)
&\propto& \sum_{k\ge 1}{1\over{\epsilon_{k} -\epsilon_{0_{+}} \pm \omega}}\,
 \Gamma^{\chi}_{k, 0_{+}}({\bf p}) \nonumber
\\ 
&-&\sum_{k\ge 1}{1\over{\epsilon_{0_{+}} -\epsilon_{-k} \pm \omega}}\,
 \Gamma^{\chi}_{0_{+}, -k}({\bf p}), \ \ 
\end{eqnarray}
which is easily seen to vanish for zero mass-gap $m\rightarrow 0$.
This shows that the  $\nu=0$ and $\nu=\pm 2$ states remain indistinguishable 
in density response even at the RPA level, 
indicating the robustness of the zero modes against perturbations.  
In particular, the static structure factor
$s({\bf p}) \sim  \int d \omega\, P_{\rm RPA}({\bf p},\omega)$ and the dielectric function
$\epsilon ({\bf p}, \omega)  = 1 - v_{\bf p}\, P({\bf p}, \omega)$
remain the same for $\nu=0$ and $\nu=\pm 2$.

\section{Summary and discussion}

In this paper we have studied the static structure factor for graphene 
in a strong magnetic field, with emphasis on revealing possible quantum signatures
that distinguish graphene from conventional QH systems.
In particular, for graphene the vacuum state is a dielectric medium 
with full of  virtual particle-hole pairs.
In other words, $\rho |\nu=0\rangle \not=0$ for graphene
while $\rho\, |\nu=0\rangle =0$ for standard QH systems.
Correspondingly the graphene vacuum state has 
a nonzero density spectral weight $\langle \rho_{\bf -p} \rho_{\bf p}\rangle|_{\nu=0}$, 
which actually diverges with the (cutoff) depth of the Dirac sea 
and generally rises with $|{\bf p}|$ significantly.
Experimentally the nonzero vacuum weight 
$\langle \rho_{\bf -p} \rho_{\bf p}\rangle|_{\nu=0}$ itself 
as well as its rise with $|{\bf p}|$, 
measured via inelastic light scattering,  
would be a clear signal of the quantum nature 
of the graphene vacuum state.

For graphene the static structure factor $s({\bf p})$ grows 
with $|{\bf p}|$ for all $\nu$, in sharp contrast to 
the behavior $s({\bf p})\rightarrow 1$ as $|{\bf p}|\rightarrow \infty$ 
of conventional QH systems; this is because of pair creation at short distances.
Actually, even the transitions among positive-energy states fail to recover 
the standard behavior and lead to $s^{(+)}({\bf p})\rightarrow 1/2$ 
for $|{\bf p}| \rightarrow \infty$, as noted  in Eq.~(\ref{splusp}).
This feature of $s^{(+)}({\bf p})$ itself, unfortunately, is not directly observable.
Nevertheless, for large filling factor $\nu$, it competes with the vacuum polarization effect 
and would make the observable subtracted structure factor $\triangle s({\bf p})$
(which is sensitive to quantum fluctuations of cutoff-independent longer wavelengths) 
saturate around 1/4 over a certain broad range of $|{\bf p}|$; 
this could be a possible signature of the underlying dynamics. 
Such unusual features of particle correlations at short distances 
are a "relativistic" effect coming from massless Dirac particles in graphene.

Of special interest, in addition, are some peculiar features of 
the lowest ($n=0_{\pm}$) Landau levels of zero energy.  
Remarkably they hardly affect the density response.
The virtual transitions, both {\em allowed} and {\em suppressed} anew 
by the presence (or absence) of the $n=0_{\pm}$ levels, combine to leave 
the density response unchanged, and this compensation persists 
even when the Coulomb interaction is taken into account in the RPA.
The graphene vacuum ($\nu=0$) state and 
the $\nu=\pm 2$ states would thus be indistinguishable 
in spectral weight $\langle \rho_{\bf -p} \rho_{\bf p}\rangle
\sim s({\bf p})$ as well as in electric susceptibility, 
although they are distinct in Hall conductance.
Experimentally one would expect a definite nonzero signal for 
 $\langle \rho_{\bf -p} \rho_{\bf p}\rangle$ 
which largely remains  the same over the range $|\nu| \le 2$,
except for some values of $\nu$ where nontrivial dynamics 
such as the fractional quantum Hall effect may come into play.

\acknowledgments

The author wishes to thank T. Morinari for useful discussion. 
This work was supported in part by a Grant-in-Aid for Scientific Research
from the Ministry of Education Science Sports and Culture of Japan 
(Grant No. 17540253).

%\newpage

%%%%%%%%%%%%%%%%%%%%% References %%%%%%%%%%%%%%%%%%%%%

\end{document}